\documentclass[aps,pra,superscriptaddress,amsmath,amssymb,preprintnumbers,floatfix,showpacs,11pt]{revtex4}
\usepackage{amssymb}
\usepackage{epsfig}
\begin{document}
\title{ Quantum properties of a superposition of squeezed displaced
two-mode vacuum and single-photon states}
\author{ Faisal A. A. El-Orany }
\email{el_orany@hotmail.com}
 \affiliation{Department of Mathematics  and computer Science,
Faculty of Science, Suez Canal University 41522,
 Ismailia, Egypt}
 \affiliation{Cyberspace Security
Laboratory, MIMOS Berhad, Technology Park Malaysia, 57000 Kuala
Lumpur, Malaysia}

\author{A-S.  F. Obada  }
 \affiliation{
 Mathematics Department, Faculty of Science, Al-Azhar
University, Nasr City,11884 Cairo,  Egypt}
\author{  Zafer M.
Asker  }
 \affiliation{Department of Mathematics,
Faculty of Science, Suez Canal University,
 Suez, Egypt}
\email{zafer.asker@hotmail.com}
\author{J. Pe\v{r}ina}
\affiliation{Department of Optics and
Joint Laboratory of Optics, \\ Palack\'y University
and Institute of Physics,  Academy of Sciences of the Czech
Republic, 17.~listopadu~50, 772 07~Olomouc, Czech Republic.}

\date{\today}

\begin{abstract}
In this paper, we study some quantum properties of a superposition
of displaced squeezed  two-mode vacuum and single-photon states,
 such as the second-order correlation
function,  the Cauchy-Schwartz inequality, quadrature squeezing,
quasiprobability distribution functions and purity. This type of
states includes two mechanisms, namely, interference in phase
space and entanglement.  We show that these states can exhibit
sub-Piossonian statistics, squeezing and deviate from the
classical Cauchy-Schwartz inequality. Moreover, the amount of
entanglement in the system can be increased by increasing the
squeezing mechanism. In the framework of the quasiprobability
distribution functions we show that  the single-mode state can
tend to thermal  state  based on the correlation mechanism.
Generation scheme for such states is given.

\end{abstract}

 \pacs{42.50.Dv,42.50.-p} \maketitle

\section{Introduction}

Developing new  states beside the traditional ones is an important
topic in quantum optics and quantum information theories. Fock
state $|n\rangle$ and coherent state $|\alpha\rangle$ are the most
commonly used states in these theories. The single-mode squeezed
states of electromagnetic field are purely quantum states since
they have less uncertainty in one quadrature than the vacuum noise
level. Additionally, these states  exhibit  a variety of
nonclassical effects, e.g. sub-Poissonian statistics \cite{har}
and oscillatory behavior in the photon-number distribution
\cite{pho}. These states can be generated via a degenerate
parametric amplifier \cite{wu}. The third type of states given in
the literature is   the two-mode squeezed states \cite{two}, which
contain quantum correlations between two different modes of the
field. The importance of these states comes from their connection
to the two-photon nonlinear processes, e.g. non-degenerate
parametric amplifier \cite{dege}. These states have been used in
the continuous-variable teleportation \cite{[7]}, quantum key
distribution \cite{[9]}, verification of EPR correlations
\cite{[12]}, etc. The single-mode state--obtained from the
two-mode squeezed state  by tracing out the other mode--cannot
exhibit squeezing \cite{bern, rabha1}. Precisely,  because of the
correlation between modes in the two-mode squeezed operator,  the
squeezing of the quantum fluctuations does not occur in the
individual modes but it occurs in the superposition of the two
modes.

A great attention has been devoted to produce
 mesoscopic superposition states. These states have interesting
distinct characteristics than the classical ones such as
interference in phase space, squeezing and quantum entanglement
\cite{quan1}. These remarkable properties present the mesoscopic
superposition states as powerful tools in quantum information
processing, metrology \cite{raimond,Glancy} and experimental studies
of decoherence \cite{zurek}. The most famous superimposed state in
the literature is the Schr\"{o}dinger-cat state \cite{sch1}.
 There are several
proposals for generating superposition of optical coherent states
in the literature. For recent review the reader can consult
\cite{Glancy} and the references cited therein. Besides  the
Schr\"{o}dinger-cat states   various types of superposition have
been developed, e.g. the superposition of squeezed and displaced
number states without \cite{oba1} and with thermal noise
\cite{elo1}. Moreover, the superposition of multiple mesoscopic
states is given in \cite{agar}, and has been generated using
resonant interaction between atoms and the field in a high quality
cavity. The superposition of the two-mode states is discussed in
\cite{chin}. These states--under certain conditions become very
close to the well-known Bell-states \cite{quan1} and  they can be
generated by a resonant bichromatic excitation of $N$ trapped ions
\cite{solano}. The entanglement of a superposition of two
bipartite states in terms of the correlation of the two states
constituting the superposition has been discussed in \cite{gilad}.

Developing  new states is an important topic for understanding the
boundary between the classical and quantum mechanics as well as to
cover the needs of the progress in the quantum information theory.
Moreover, the investigation  of nonclassical effects of the quantum
states is of considerable and continuing interest, since it plays an
important role both fundamentally and practically in the quantum
information theory. Throughout this paper we study the quantum
properties of the superposition of squeezed displaced two-mode
number states (STDSN), in particular, the vacuum and single-photon
states. In these states the squeezing mechanism is involved via
non-degenerate squeezed operator. These states are different from
the superposition  of the single-mode states \cite{oba1,elo1} in the
following sense: They include two mechanisms: (i) entanglement
and/or correlation between the two-modes. (ii) Two-mode
interferences in phase space. These states can be generated via
two-mode trapped ions \cite{trapa}, as we will show in section VI.
For STDSN  we study the single-mode second-order correlation,
Cauchy-Schwartz inequality, quadrature squeezing, quasiprobability
functions and purity. We show that the nonclassical effects are
remarkable in the different quantities. Also
  the single-mode state  tends to the  thermal state based on the correlation
mechanism and the amount of entanglement  can be increased by
increasing the squeezing mechanism.

We perform this investigation in the following order. In section
\ref{S:sec2}  we introduce the state formalism and comment on its
photon-number distribution. In section \ref{S:sec3} we discuss the
second-order correlation function and Cauchy-Schwartz inequality. In
section \ref{S:sec4} the quadrature squeezing in the framework of
principal squeezing is  investigated.
 In section \ref{S:sec5} quasiprobability distribution functions
 and the purity are investigated. The generation of the STDSN is  discussed
in section \ref{S:sec6}, however, the conclusions are summarized
in section \ref{S:sec7}.

\section{State formalism}\label{S:sec2}
The correlated  two-mode squeezed states are connected with the
two-mode squeeze operator, which has the form:
\begin{equation}
\hat{S}(r)=\exp[\frac{r}{2}(\hat{a}\hat{b}-\hat{a}^{\dag}\hat{b}^{\dag})],\label{sec1}
\end{equation}
where $\hat{a}\quad (\hat{a}^\dagger)$ and $\hat{b}\quad
(\hat{b}^\dagger)$ denote the annihilation (creation) operators of
the first (signal) and second (idler) mode, respectively. By means
of this operator and  the superposition principle we develop a new
class of states, namely,  superposition of squeezed displaced
two-mode number states (STDSN) as:
\begin{equation}
|\psi\rangle=|r,\alpha,\beta\rangle_{\epsilon}=\lambda_{\epsilon}[\hat{D}(\alpha_{1},\alpha_{2})
+\epsilon\hat{D}(-\alpha_{1},-\alpha_{2})]\hat{S}(r)|n,m\rangle,
\label{sec2}
\end{equation}
where $\epsilon=|\epsilon|\exp(i\phi),\hat{S}(r)$ is given by
(\ref{sec1}) and
 $ \hat{D}(\alpha_{1},\alpha_{2})$ is the two-mode displaced
 operator defined  as:
\begin{equation}
\hat{D}(\alpha_{1},\alpha_{2})=\hat{D}(\alpha_{1})\hat{D}(\alpha_{2})=
\exp(\alpha_1\hat{a}^{\dag}-\alpha_1^{*}\hat{a})
\exp(\alpha_2\hat{b}^{\dag}-\alpha_2^{*}\hat{b}) \label{sec3}
\end{equation}
and $\alpha_j$ is generally a complex parameter (a field
amplitude), however, throughout the investigation in this paper it
will be considered real. Also the prefactor $\lambda_{\epsilon}$
is the normalization constant, which can be easily evaluated as:
\begin{equation}
|\lambda_{\epsilon}|^{-2}=1+|\epsilon|^{2}+2|\epsilon|\mu{\rm
L}_n(4t_1^2){\rm L}_m(4t_2^2)\cos\phi, \label{sec4}
\end{equation}
where
\begin{equation}
 t_{1}=\alpha_{1}C_r+\alpha_{2}S_r, \quad
 t_{2}=\alpha_{2}C_r+\alpha_{1}S_r,\quad  S_r=\sinh r, C_r=\cosh r,\quad
 \mu=\exp[-2(t^{2}_{1}+t^{2}_{2})]
\label{sec5}
\end{equation}
and ${\rm L}_n(.)$ is the Laguerre polynomial of order $n$ (see
(8) below).
 Throughout the paper we
study only two choices for the parameter $|\epsilon|$, namely, $1$
and $0$, however, for the parameter $\phi$ we take the values
$0,\pi$ and $\pi/2$. Precisely, when $|\epsilon|=1$ and
$\phi=0,\pi,\pi/2$  the states (\ref{sec2}) are called even-type,
odd-type and Yurke-type states, respectively.

When $|n,m\rangle=|0,0\rangle$ the states (\ref{sec2}) can be
expressed in a closed form in terms of the Fock states
 \cite{cave} as:
\begin{equation}
|\psi\rangle=\sum\limits_{n_1,n_2=0}^{\infty}
C(n_1,n_2)|n_1,n_2\rangle,  \label{sec6}
\end{equation}
where
\begin{eqnarray}
\begin{array}{lr}
C(n_1,n_2)=\lambda_{\epsilon}[1+(-1)^{n_1+n_2}\epsilon]\frac{1}{\cosh
r}
\exp[-\frac{1}{2}(\alpha_1\mu_1+\alpha_2\mu_2)]\sqrt{\frac{M!}{N!}}
(\mu_1)^{n_1-M}(\mu_2)^{n_2-M}\\
\\ \times (\tanh r)^M {\rm L_M^{N-M}}\left(-\frac{\mu_1\mu_2}{\tanh
r}\right),\\
\\
\mu_1=\alpha_1-\alpha_2 \tanh r, \quad \mu_2=\alpha_2-\alpha_1
\tanh r, \quad M=min(n_1,n_2),\quad N=max(n_1,n_2)
 \label{sec7}
 \end{array}
\end{eqnarray}
and ${\rm L}_k^{\upsilon}(.)$ is   the associated Laguerre
polynomial having the form:
\begin{equation}\label{reply1}
    {\rm L}_k^{\upsilon}(x)=\sum\limits_{l=0}^{k}
    \frac{(\upsilon+k)!(-x)^l}{(\upsilon+l)!(k-l)!l!}.
\end{equation}
The photon-number distribution of (\ref{sec6}) can be evaluated
as:
\begin{equation}
P(m_1,m_2)=|C(m_1,m_2)|^2 , \label{sec8}
 \end{equation}
where $C(m_1,m_2)$ is given by (\ref{sec7}). It is obvious that
$P(m_1,m_2)$ can exhibit pairwise oscillations based on the values
of the sum $m_1+m_2$, even if $r=0$. We have to remark that the
components of the STDSN can exhibit oscillatory behavior in
$P(m_1,m_2)$ \cite{cave}, apart from the superposition mechanism,
which can make this behavior more or less pronounced.  Moreover, the
single-mode photon-number distribution can be obtained via the
relation:
\begin{equation}
P(m_1)=\sum\limits_{m_2=0}^{\infty}|C(m_1,m_2)|^2. \label{sec9}
 \end{equation}
In $P(m_1)$ the occurrence of the oscillatory behavior results
from the interference mechanism. We can explain this fact for the
simplest case $r=0,n=m=0, \epsilon=\exp(i\phi)$ and hence
(\ref{sec9}) reduces to
\begin{equation}
P(m_1)=2|\lambda_\epsilon|^2
\exp(-\alpha_1^2)\frac{\alpha_1^{2m_1}}{m_1!}[1+(-1)^{m_1}
\exp(-2\alpha_2^2)\cos\phi]. \label{sec10}
 \end{equation}
 The oscillatory behavior in $P(m_1)$ depends on the values of
 $\alpha_2$ and $\phi$, i.e. for large values of  $\alpha_2$,
 $P(m_1)$ tends to that of the coherent state. This means
 that one can use the second mode to control the nonclassical
 effects in the first mode and vice versa.

In the following sections we investigate the properties of the
state (\ref{sec2}). For the sake of simplicity we treat the
second-order correlation function, Cauchy-Schwartz inequality,
squeezing and purity using the form (\ref{sec6}) (,i.e.,
$|n,m\rangle=|0,0\rangle$), however, the quasiprobability
functions are given for the case $|n,m\rangle=|0,1\rangle$. This
is to estimate a global information on the generic form.

\section{Second-order correlation function and  Cauchy-Schwartz inequality}\label{S:sec3}
In this section we  investigate the behavior of  the second-order
correlation function and Cauchy-Schwartz inequality for the state
(\ref{sec6}). These two quantities can give information on the
correlation between the modes in the quantum system. The
second-order correlation function for the first mode, e.g.
$\hat{a}$, is defined by

\begin{equation}\label{pw3}
g^{(2)}(0)=\frac{\langle\hat{a}^{\dagger 2}\hat{a}^{
2}\rangle}{\langle \hat{a}^{\dagger}\hat{a}\rangle^2}-1,
\end{equation}
where $g^{(2)}(0)=0$ for Poissonian statistics (standard case),
$g^{(2)}(0)<0$ for sub-Poissonian statistics (nonclassical
effects) and $g^{(2)}(0)>0$ for super-Poissonian statistics
(classical effects). The second-order correlation function can be
measured by a set of two detectors \cite{mandl}, e.g. the standard
Hanbury Brown-Twiss coincidence arrangement. For this quantity we
restrict the discussion to the first-mode only.
 For this mode one can easily obtain:

\begin{equation}
\langle
\hat{a}^{\dag}\hat{a}\rangle=|\lambda_{\epsilon}|^{2}\{(S_r^{2}+\alpha_{1}^{2})(1+|\epsilon|^{2})+
2|\epsilon|\mu \cos\phi\left[S^{2}_r-4t_{1}t_{2}S_rC_r
+2\alpha_{1}t_{2}S_r-2\alpha_{1}t_{1}C_r+\alpha_{1}^{2}\right]\},
\label{secc2}
\end{equation}
\begin{eqnarray}
\begin{array}{lr}
\langle
\hat{a}^{\dag2}\hat{a}^{2}\rangle=|\lambda_{\epsilon}|^{2}\{(\alpha_{1}^{4}+2S^{4}_r+4\alpha_{1}^{2}S^{2}_r)
(1+|\epsilon|^{2})+2|\epsilon|\mu \cos\phi
\left[4\alpha^{2}_{1}C_r^{2}t^{2}_{1}+16S_r^{2}C_r^{2}t^{2}_{1}t^{2}_{2}\right.
\\
+16\alpha_{1}S_rC_r^{2}t^{2}_{1}t_{2}+\alpha^{4}_{1}+4\alpha^{2}_{1}S_r^{2}t^{2}_{2}+2S_r^{4}
+8\alpha_{1}S_r^{3}t_{2}+4\alpha^{3}_{1}S_rt_{2}-8\alpha^{2}_{1}S_r^{2}t_2-4\alpha^{3}_{1}C_rt_{1}
\\\left.
-16\alpha_{1}S_r^{2}C_rt_{1}t^{2}_{2}-16S_r^{3}C_rt_{1}t_{2}-16\alpha^{2}_{1}S_rC_r
t_{1}t_{2}-8\alpha_{1}S_r^{2}C_rt_{1}\right]\}. \label{secc3}
 \end{array}
\end{eqnarray}
It is worth mentioning that the most general cases for
 equations (\ref{secc2}) and (\ref{secc3}) have been given in \cite{mank}
 for the  multimode squeezed cat
states but with different parameterizations.

 Substituting  (\ref{secc2}), (\ref{secc3}) into
(\ref{pw3}) and taking  $r=0$ we obtain

\begin{equation}
g^{(2)}(0)=\frac{[1+|\epsilon|^{2}+2|\epsilon|\exp[-2(\alpha_{1}^{2}+\alpha_{2}^{2})]\cos\phi]
^{2}}{[1+|\epsilon|^{2}-2|\epsilon|\exp[-2(\alpha_{1}^{2}+\alpha_{2}^{2})]\cos\phi]^{2}}-1.
\label{secc4}
\end{equation}
\begin{figure}
    \includegraphics[width=1.0\linewidth]{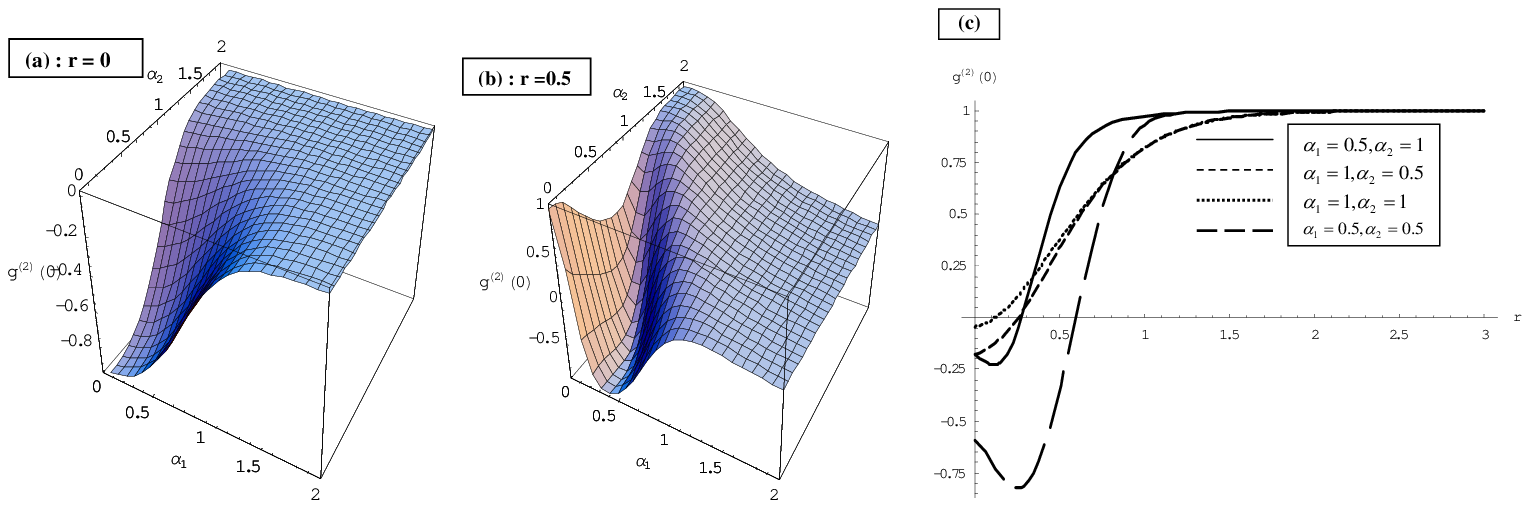}
\caption{ The second-order correlation function of the first mode
 for $(|\epsilon|,\phi)=(1,\pi)$ against $(\alpha_1,\alpha_2)$
(a)--(b) and $r$ (c) for different values of the parameters as
indicated. }
\end{figure}

From (\ref{secc4}) it is obvious  that the sub-Poissonian statistics
can occur only for $\phi=\pi$  and
$2(\alpha_{1}^{2}+\alpha_{2}^{2})$ being small. This means that the
odd-type state can exhibit nonclassical effects in the framework of
 $g^{(2)}(0)$. In this case, the mode under consideration reduces
to the standard odd-coherent state with the components
$|\pm\sqrt{\alpha_1^2+\alpha_2^2}\rangle$.  The obvious remark is:
when the mode $\hat{a}$ is prepared in the vacuum state
$|0\rangle$, its $g^{(2)}(0)$  can exhibit sub-Poissonian
statistics based on the values of $\alpha_2$ of the second mode.
Similar argument can be given to the second mode. This reflects
the role of the correlation between the modes in the system, which
leads to the possibilities of controlling one mode by the other
one. Now we draw the attention to the general case when the
squeezing mechanism is involved. We have noted that the even-type
 and the Yurke-type states cannot exhibit
sub-Poissonian statistics. Information about $g^{(2)}(0)$ of the
odd-type states is depicted in Figs. 1(a)--(c) for given values of
the parameters $\alpha_{1}$ and $\alpha_{2}$. From Fig. 1(a) one
can observe the occurrence of the sub-Poissonian statistics, in
particular, for small values of $\alpha^{,}s$.  When the squeezing
mechanism is involved, the amounts of the nonclassical effects in
$g^{(2)}(0)$ decrease and eventually vanish for large value of $r$
(see Fig. 1(b)).  For $\alpha_1=\alpha_2=0$ and $r\neq 0$ state
(\ref{sec6}) reduces to the two-mode squeezed vacuum state. In
this case we have $g^{(2)}(0)=1$, which is independent of $r$.
This is remarkable in Fig. 1(b), which shows  sub-Poissonian
statistics only when one or both of $\alpha_j>0$. Fig. 1(c) gives
the range of the parameter $r$ (for certain values of $\alpha_j$)
for which the sub-Piossonian statistics occur. It is obvious that
the smaller the values of $\alpha_j$, the greater this range is.

\begin{figure}
    \includegraphics[width=1.0\linewidth]{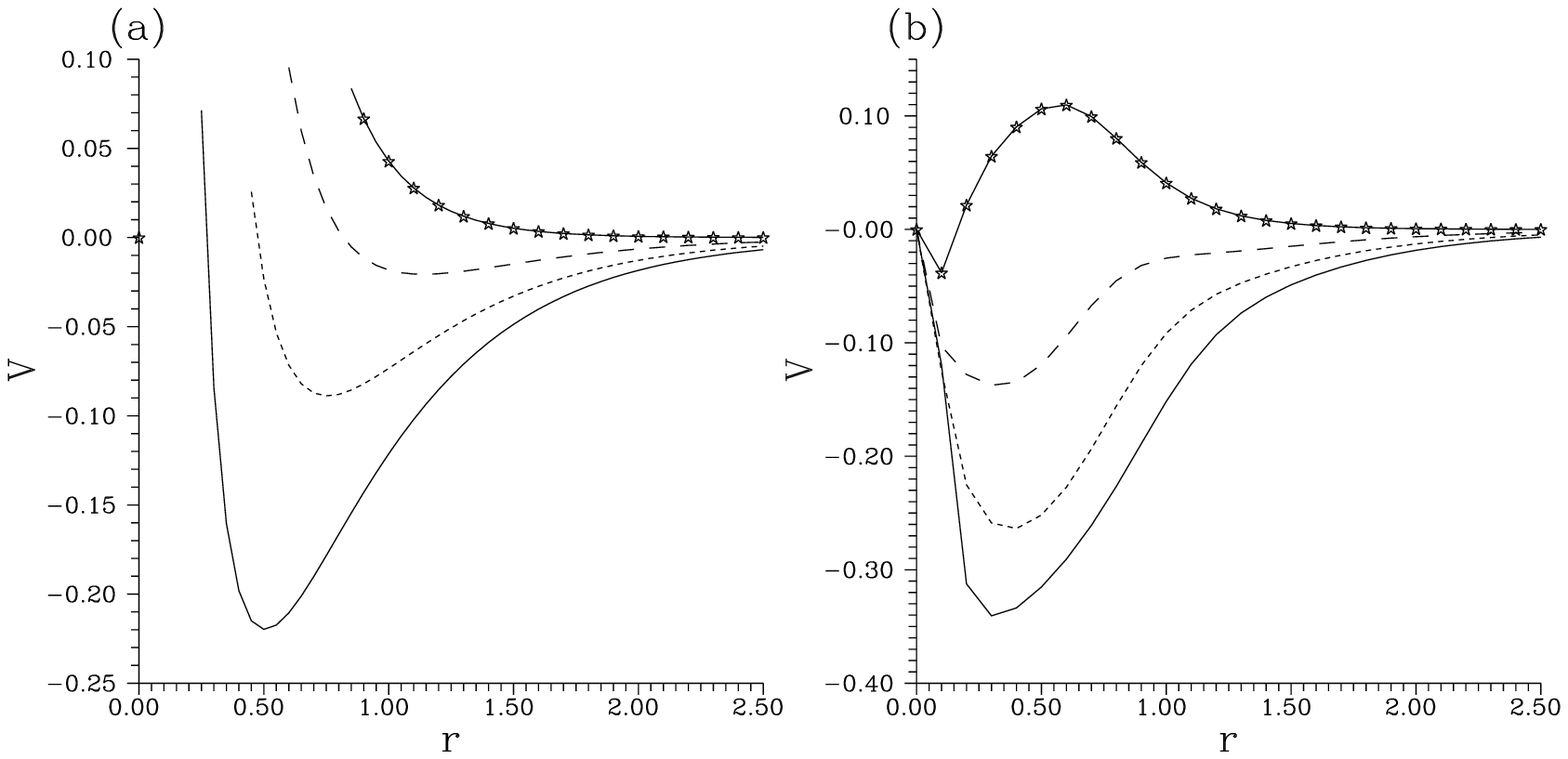}
\caption{ The factor $V$ against the parameter $r$ for
$(|\epsilon|,\phi,\alpha_1)=(1,\pi/2,0.5)$ (a) and $(1,\pi,0.5)$
(b) for $\alpha_2=0.2$ (solid), $0.3$ (short-dashed curve), $0.4$
(long-dashed curve) and $0.5$ (star-centered curve).  }
\end{figure}

We conclude this section by investigating the  intermodal
correlations in terms of the deviation from the classical
Cauchy-Schwarz inequality. Classically, Cauchy-Schwarz inequality
has the form \cite{[32]}:
\begin{equation}\label{reply10}
    \langle I_1I_2\rangle\leq \langle I_1\rangle\langle
    I_2\rangle,
\end{equation}
 where $I_j, j = 1, 2$ are classical
intensities of light measured by different detectors in a
double-beam experiment. In quantum theory, the deviation from this
classical inequality can be represented as $V<0$, where the factor
$V$ takes the form \cite{agrw}:
\begin{equation}
V=\frac{\sqrt{\langle \hat{a}^{\dag 2}\hat{a}^2\rangle\langle
\hat{b}^{\dag 2}\hat{b}^2\rangle}}{\langle
\hat{a}^{\dag}\hat{b}^{\dag}\hat{a}\hat{b}\rangle}-1.
\label{secc5}
\end{equation}
Occurrence of  negative values in  $V$ means that the intermodal
correlation is larger than the correlation between photons in the
same mode \cite{kinn} and this indicates strong deviation from the
classical Cauchy-Schwartz inequality.  The origin in this
deviation is that in the quantum mechanical treatment we involve
pseudodistributions instead of the true ones. This implies that
the Glauber-Sudarshan $P$-function possesses strong quantum
properties \cite{agrw}. Moreover, the deviation from the
Cauchy-Schwarz inequality can be observed in a two-photon
interference experiment \cite{mand}.  For completeness, the
expectation value $\langle
\hat{a}^{\dag}\hat{b}^{\dag}\hat{a}\hat{b}\rangle$
 for the state (\ref{sec6}) can be easily evaluated as:

\begin{eqnarray}
\begin{array}{lr}
\langle
\hat{a}^{\dag}\hat{a}\hat{b}^{\dag}\hat{b}\rangle=|\lambda_\epsilon|^{2}
\left\{[S^{4}_r+(S_rC_r-\alpha_{1}\alpha_{2})^2+(\alpha_{1}^{2}+\alpha_{2}^{2})S^{2}_r
](1+|\epsilon|^{2})\right.
\\
\\
+2|\epsilon|\mu\cos\phi\left[ S^{2}_ r\cosh(2r)
-\alpha_{1}\alpha_{2}\sinh(2r) +(\alpha_{2}^2+\alpha_{1}^2)S_r^2
\right.\\
\\
-2t_{1}t_{2}[\sinh(2r)(2S^{2}_r+\alpha^{2}_{1}+\alpha^{2}_{2}+\cosh(2r))-2\alpha_{1}\alpha_{2}
\cosh(2r)]+4\sinh^2(2r)t^{2}_{1}t^{2}_{2}
\\
\\
+2t_{2}[-S_r\sinh(2r)\alpha_{2}+\alpha_{1}S_r\cosh(2r)-\alpha_{1}\alpha_{2}(\alpha_{1}C_r-\alpha_{2}S_r)]
\\
\\
-2t_{1}[(\alpha_{1}C_r-\alpha_{2}S_r)S^{2}_r+(\alpha_{1}S_r-\alpha_{2}C_r)(S_rC_r
-\alpha_{1}\alpha_{2})]
\\
\\
-\left.
4\sinh(2r)(\alpha_{1}C_r-\alpha_{2}S_r)(t_{2}-t_{1})t_{1}t_{2}-
2\alpha_{1}\alpha_{2}\sinh(2r)(t^{2}_{1}+t^{2}_{2})+
\alpha^{2}_{1}\alpha^{2}_{2}\right\}.
\label{secc6}
 \end{array}
\end{eqnarray}
The expectation value $\langle \hat{b}^{\dag 2}\hat{b}^{2}\rangle$
can be obtained from (\ref{secc3}) using the interchange
$\alpha_1\longleftrightarrow \alpha_2$. One can easily find $V=0$
for $r=0$.  Generally, we have noted that  $V<0$  only when
$\alpha_j$ are small (see Figs. 2). Fig. 2(a) is given for  the
Yurke-type state which is  identical to that of the two-mode
squeezed displaced states (cf. (\ref{secc3}) and (\ref{secc6}) for
$\phi=\pi/2$). From these figures the strongest deviation from the
classical
 inequality occurs for $\alpha_j(\neq 0)$ and $r$  small, i.e.
the photons are more strongly correlated than it is classically
possible, and then the curve monotonically increases as $r$
increases. As the values of $\alpha_2$ increase the negative
values in the factor $V$ decrease and eventually disappear
(compare different curves in these figures). Also a comparison
between Fig. 2(a) and Fig. 2(b) shows that the nonclassical
effects occurred in the factor $V$ for the odd-type state  are
greater than those in the state with Yurke-type state.

\section{Quadrature squeezing}\label{S:sec4}
In this section we discuss the quadrature squeezing for the state
under consideration. As it is well known, the quadrature squeezing
can be measured by a homodyne detector in which the signal is
superimposed on a strong coherent beam of the local oscillator
\cite{homo}. Here we use the notion of the principal squeezing
\cite{princ}, which can give one form for the single-mode and
two-mode cases. In this respect, we define the two quadratures in
the following forms:
 \begin{equation}\label{sect1}
\hat{X}=\hat{X}_1 \cos\nu + \hat{X}_2 \sin\nu,\quad \hat{Y}=
\hat{Y}_1 \cos\nu + \hat{Y}_2 \sin\nu,
\end{equation}
where the subscripts $1$ and $2$ stand for the first and second
mode, respectively, and $\nu$ is a rotation angle. When
$\nu=0,\pi/2,\pi/4$ the quadratures (\ref{sect1}) yield those of
the first mode, second mode and compound modes, respectively. For
the first mode the quadrature operators can be defined as:
\begin{equation}\label{reply21}
\hat{X}_1=\frac{1}{2}\left(\hat{a}+\hat{a}^{\dag}\right),\quad
\hat{Y}_1=\frac{1}{2i}\left(\hat{a}-\hat{a}^{\dag}\right).
\end{equation}
Similar definition can be quoted for the second mode via the
interchange $\hat{a}\longrightarrow\hat{b}$. The quadratures
(\ref{sect1}) satisfy the following commutation rule:

 \begin{equation}\label{sect2}
\left[\hat{X}, \hat{Y}\right]=\frac{i}{2}.
\end{equation}
Therefore, the squeezing factors associated with the $\hat{X}$ and
$\hat{Y}$ can be expressed as:
\begin{figure}
    \includegraphics[width=1.0\linewidth]{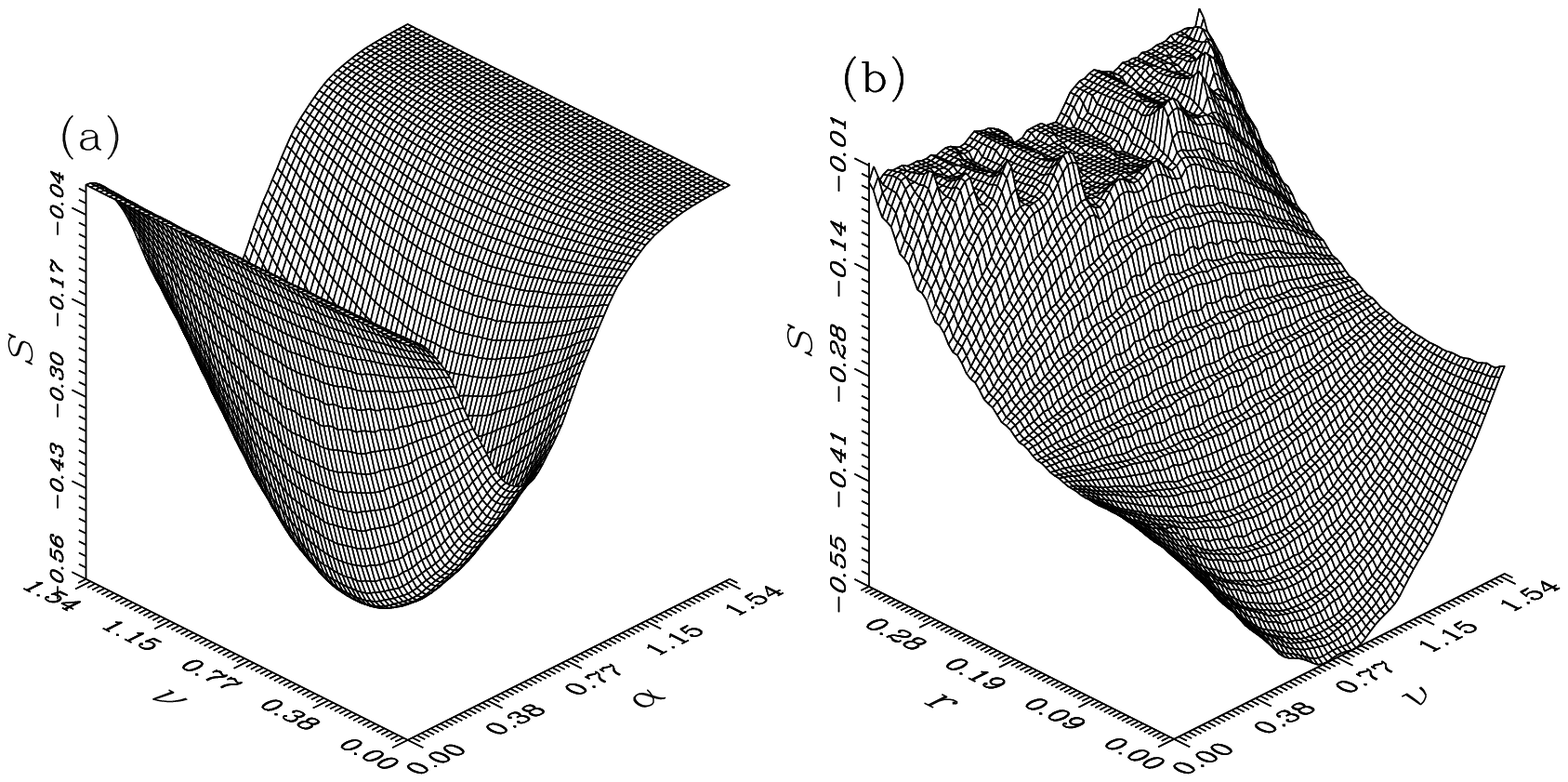}
\caption{ Squeezing factor    $S$ for $(|\epsilon|,\phi)=(1,0)$
with $(r,\alpha_1,\alpha_2)=(0,\alpha,\alpha)$ (a) and
$\alpha_1=\alpha_2=0.6$ (b).
 }
\end{figure}
 \begin{eqnarray}
 \begin{array}{lr}
F=\langle(\vartriangle \hat{X})^2\rangle-1 \\
\\
= F_1\cos^2\nu + F_2\sin^2\nu +
F_{c}\sin(2\nu),\\
\\
S= \langle(\vartriangle \hat{Y})^2\rangle-1\\
\\
= S_1\cos^2\nu + S_2\sin^2\nu + S_{c}\sin(2\nu). \label{sect10}
\end{array}
\end{eqnarray}
where $F_j,S_j, j=1,2,c$, take the forms:
 \begin{eqnarray}
 \begin{array}{lr}
F_1 =2S_r^2+
4\alpha_1^2|\lambda_\epsilon|^2[1+|\epsilon|^2-|\lambda_\epsilon|^2
(1-|\epsilon|^2)^2],\\
\\
S_1=2S_r^2-
4|\lambda_\epsilon|^2|\epsilon|\mu(\alpha_1^2+4t_1t_2S_rC_r)\cos\phi\\
\\
- 4|\lambda_\epsilon|^2|\epsilon|\mu
(2t_1C_r-\alpha_1)^2\left(\cos\phi+4|\lambda_\epsilon|^2|\epsilon|^2\mu\sin^2\phi\right),\\
\\
F_{c}=4\alpha_1\alpha_2-2S_rC_r-8\alpha_1\alpha_2 |\epsilon|\mu
|\lambda_\epsilon|^2\cos\phi-4|\lambda_\epsilon|^2(1-|\epsilon|^2)^2\alpha_1\alpha_2,\\
\\
S_{c}= 2C_rS_r-
8|\epsilon||\lambda_\epsilon|^2\mu(2t_1C_r-\alpha_1)(t_1S_r+t_2C_r)\cos\phi
- 16|\epsilon|^2|\lambda_\epsilon|^4\mu^2
(\alpha_1-2t_1C_r)(\alpha_2-2t_2C_r)\sin^2\phi. \label{sectr10}
\end{array}
\end{eqnarray}
 The expressions for $F_2, S_2$
 can be obtained from $F_1, S_1$ via the
interchange $(\alpha_1,t_1)\longleftrightarrow (\alpha_2,t_2)$.
 The system is said to
be squeezed in $x$-quadrature or $y$-quadrature if $F<0$ or $S<0$,
respectively. When $|\epsilon|=0$ or $\alpha_1=\alpha_2=0$ the
expressions (\ref{sect10}) reduce to
\begin{equation}
F= 2S_r[S_r-C_r\sin(2\nu)],\quad S= 2S_r[S_r+C_r\sin(2\nu)].
\label{reply11}
\end{equation}
It is evident that for $\nu=0$ or $\pi$, we obtain $F=S=2S^{2}_r$,
i.e. single-mode squeezing does not occur for these cases
\cite{Loudon}. On the other hand, for $\nu=\pi/4$ (, i.e. the
compound squeezing factor) we have:
\begin{equation}\label{reply12}
    F= \exp(-2r)-1,\quad S= \exp(2r)-1.
\end{equation}
Trivial remark from (\ref{reply12}), squeezing occurs in the
$x$-component only \cite{Loudon}. This is related to  strong
 correlation between the two modes. This behavior is reversed for
 the superposition state (, i.e., $|\epsilon|\neq 0$), where
 squeezing can occur in
the $y$-component for certain values of $\alpha^{,}s$ and $r$. For
$(|\epsilon|,r)=(1,0)$ the expressions (\ref{sect10}) reduce to:
 \begin{eqnarray}
 \begin{array}{lr}
F =8|\lambda_\epsilon|^2[\alpha_1\cos\nu+\alpha_2\sin\nu]^2,
\\
\\
S =-8|\lambda_\epsilon|^2\mu (\alpha_1\cos\nu+\alpha_2\sin\nu)^2
(\cos\phi+2|\lambda_\epsilon|^2\mu \sin^2\phi).
 \label{reply13}
\end{array}
\end{eqnarray}
It is evident that  squeezing can occur only in the $y$-component
for the cases $(|\epsilon|,\phi)=(1,0),(1,\pi/2)$. Moreover,  to
obtain squeezing from the single-mode case, the mode under
consideration should be prepared in a state different from vacuum.
When the values of $\alpha_j$ increase,  i.e. the correlation
between the two modes starts to play a role, the coefficient $\mu$
goes rapidly to zero decreasing  squeezing inherited  in the
system.
  This shows how one can control
the behavior of one of the modes by the other one. In Figs. 3 we
give information on  the even-type states  for the given values of
the parameters. In these figures we consider $0\leq\nu\leq \pi/2$,
which is sufficient to obtain  full information about the
different quadratures. Additionally, the behavior of the
quadratures in the range $\pi/2\leq \nu\leq \pi$ is just a mirror
image to that in $0\leq\nu\leq \pi/2$. In figure (a) we take
$\alpha_1=\alpha_2=\alpha$ and $r=0$. From this figure--regardless
of the values of $\nu$--squeezing occurs within the range
$0<\alpha\leq 1.5$ otherwise $S\geq 0$. Furthermore, the minimum
value in $S$ is observable around $\nu=\pi/4$ and $\alpha\simeq
0.6$. On the other hand, we have noted that squeezing mechanism
decreases the amount of  squeezing involved in the system. This
obvious in Fig. 3(b), which shows the range of $r$ over which
squeezing is available, i.e. $0\leq r\leq 0.35$. In other words,
$r=0.35$ is the critical value for $\alpha_1=\alpha_2=0.6$. This
critical value is $\alpha$ dependent, however, we have found
difficultly to obtain an analytical  form for it. Comparison
between Figs. 3(a) and (b) shows that involving the two mechanisms
(i.e., squeezing and superposition) in the system destroys the
nonclassical effects contributed by each one independently.

\section{Quasiprobability distribution function}\label{S:sec5}
Quasiprobability distribution functions, namely, Husimi function
($Q$), Wigner function ($W$), and Glauber $P$ functions \cite{wign},
are important tools in quantum optics. Knowing these functions, all
nonclassical effects can be predicted and the different moments of
the operators can be evaluated. Most important, these functions can
be measured by various means, e.g. photon counting experiments
\cite{con}, using simple experiments similar to that used  in the
cavity QED and ion traps \cite{ion1,ion2}, and homodyne tomography
\cite{tom}. In this section, we  investigate the single-mode
quasiprobability distribution functions, in particular, $W$ and $Q$
functions as well as the purity. We start with the symmetric
characteristic function $C_w(\beta)$ of the first mode, which is
defined as:

\begin{equation}
C_w(\beta)={\rm
Tr}[\hat{\rho}\exp(\beta\hat{a}^{\dag}-\beta^{*}\hat{a})],\label{secf1}
\end{equation}
where $\hat{\rho}$ is the density matrix of the system under
consideration. It is mentioning worth that the moments of the
bosonic operators in  symmetric form   can be evaluated from
$C_w(\beta)$ by differentiation.  From (\ref{sec2}) and
(\ref{secf1}) one can easily obtain:

\begin{eqnarray}
\begin{array}{lr}
C_{w}(\beta)=|\lambda_{\epsilon}|^{2}|\{
\exp\left(-\frac{|\beta|^2}{2}\cosh(2r)\right)\left[\exp((\beta-\beta^{*})\alpha_{1})+
|\epsilon|^{2}\exp((\beta^{*}-\beta)
\alpha_{1})\right]{\rm L}_m(S_r^2|\beta|^2){\rm L}_n(C_r^2|\beta|^2)\\
\\
+|\epsilon|\left[\exp(-i\phi-\frac{k_{+}+k'_{+}}{2}) {\rm
L}_m(k'_+){\rm L}_n(k_+) +\exp(i\phi-\frac{k_{-}+k'_{-}}{2}) {\rm
L}_m(k'_-){\rm L}_n(k_-) \right]\},\label{secf2}
\end{array}
\end{eqnarray}
where
\begin{equation}
k_{\pm}=|\beta Cr\pm 2t_{1}|^{2}, \quad k'_{\pm}=|\beta^{*} Sr\pm
2t_{2}|^{2} \label{sef3}
\end{equation}
and ${\rm L}_m(.)$ is the Lagurre polynomial, which can be
obtained from (\ref{reply1}) by simply setting $\upsilon=0$.
 The  $W$  and $Q$ functions can be
evaluated, respectively, through the following relations:
\begin{eqnarray}
\begin{array}{lr}
W(z)=\pi^{-2}\int d^{2}\beta C_{w}(\beta)\exp(z\beta^{*}-\beta
z^{*}), \\
\\
Q(z)=\pi^{-2}\int d^{2}\beta C_{w}(\beta) \exp(z\beta^{*}-\beta
z^{*}-\frac{1}{2}|\beta|^{2}), \label{secf4}
\end{array}
\end{eqnarray}
where $z=x+iy$. Generally, it is difficult to obtain closed forms
for these functions for $m\neq 0, n\neq 0$, however, the
integration can be  numerically treated.
 Therefore,
 we restrict ourselves  to the case
$n=0, m\neq 0$, which is sufficient to obtain information on the
system. On substituting (\ref{secf2}) into (\ref{secf4}) and
carrying out the integration  we arrive at:
\begin{eqnarray}
\begin{array}{lr}
W(z)=\frac{2|\lambda_{\epsilon}|^{2}}{\pi \cosh^{m+1}
(2r)}\{\exp[-\frac{2|z-\alpha_1|^2}{\cosh(2r)}] {\rm
L}_m[-\frac{4S_r^2}{\cosh(2r)}|z-\alpha_1|^2] + |\epsilon|^{2}
\exp[-\frac{2|z+\alpha_1|^2}{\cosh(2r)}] {\rm
L}_m[-\frac{4S_r^2}{\cosh(2r)}|z+\alpha_1|^2]
\\
\\
+2|\epsilon|\exp[-\frac{2}{\cosh(2r)}
(\alpha_2^2+x^{2}+y^{2})]{\rm Re}\left[\exp\left(-i\phi
+i\frac{4y\Lambda}{\cosh(2r)}\right) {\rm L}_m(h)\right] \},
\\
\\
Q(z)=\frac{|\lambda_{\epsilon}|^{2}}{\pi C^{2m+2}_r}\{
\exp[-\frac{|z-\alpha_1|^2}{C_r^2}] {\rm
L}_m[-|z-\alpha_1|^2\tanh^2 r] +|\epsilon|^{2}
\exp[-\frac{|z+\alpha_1|^2}{C_r^2}] {\rm
L}_m[-|z+\alpha_1|^2\tanh^2 r]
\\
\\
+2|\epsilon|\exp[-\frac{1}{C_r^2}(\alpha_1^2+2t^2_2+x^{2}+y^{2})]
{\rm Re}\left[\exp\left(-i\phi -i\frac{2y\Lambda}{C^2_r}\right)
{\rm L}_m(h')\right]\},\label{secf5}
\end{array}
\end{eqnarray}
where
\begin{eqnarray}
\begin{array}{lr}
\Lambda=t_{1}C_r+t_{2}S_r,\quad
h=\frac{4}{\cosh(2r)}[\alpha_2^2C^2_r+iy\alpha_2\sinh(2r)-S_r^2|z|^2],
 \\
  \\
  h'=\frac{1}{C^2_r}[(\alpha_2C_r+t_2)^2+2iyS_r
  (2\alpha_2C_r+\alpha_1S_r)-|z|^2S^2_r]. \label{scf5}
\end{array}
\end{eqnarray}
\begin{figure}
    \includegraphics[width=1.0\linewidth]{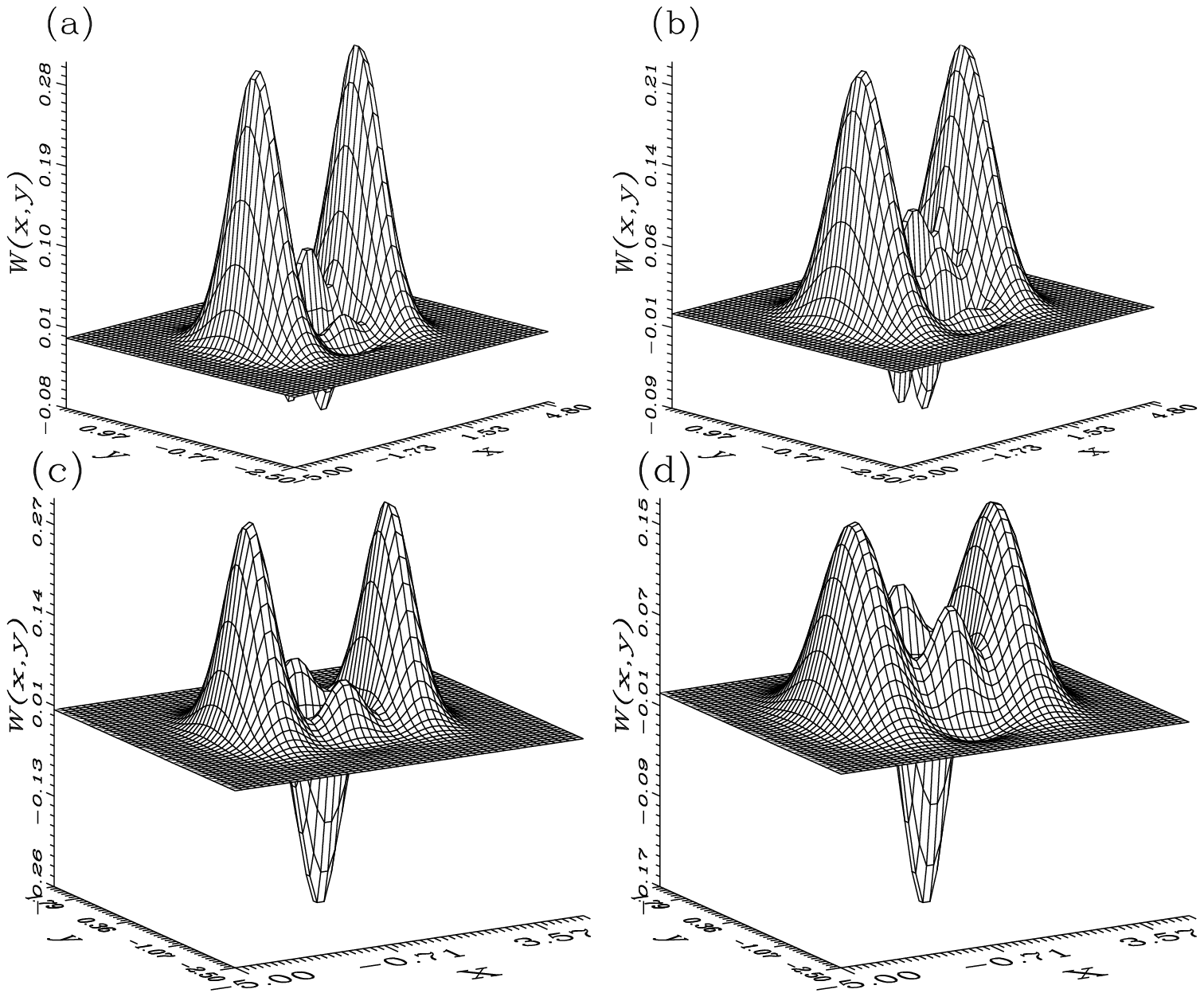}
\caption{ The $W$ function of the STDSN for
$(\alpha_1,\alpha_2,|\epsilon|,\phi)=(2,0.9,1,0)$ with
$(r,m)=(0,0)$ (a),$(0.4,0)$ (b), $(0,1)$ (c) and $(0.4,1)$ (d). }
\end{figure}
In the derivation of (\ref{secf5}) we have used the generating
function of Laguerre polynomial \cite{scfr14} namely:
\begin{equation}\label{reply30}
\frac{\exp(-\frac{ty}{1-t})}{1-t}=\sum^{\infty}_{n=0} t^{n} {\rm
L}_{n}(y),
\end{equation}
and the  following identity \cite{perin}:

\begin{eqnarray}
\begin{array}{lr}
 \int \exp
\left[-B|\beta|^{2}+(c/2)\beta^{*2}+(c_{1}/2)\beta^{2}+
\gamma_{1}\beta+\gamma\beta^{*}\right] d ^{2}\beta\\
\\
 =\frac{\pi}{\sqrt{K}} \exp
\left\{\frac{1}{K}[\gamma
\gamma_{1}B+\gamma^{2}(c_{1}/2)+\gamma^{2}_{1}(c/2)]\right\},\label{reply31}
\end{array}
\end{eqnarray}
where $K=B^2-cc_1$ if ${\rm Re}[B+\frac{1}{2}(c+c_1)]$ and ${\rm
Re}K>0$. In the  Appendix we show that the quasiprobability
functions (\ref{secf5}) are normalized. It is mentioning worth
that the explicit analytical expressions for the $W$ and $Q$
functions of the even/odd superpositions of two-mode squeezed
coherent states, which are special cases of (\ref{secf5}) by
simply setting $m=0$, were obtained earlier in \cite{manko}.   We
start the investigation with the $W$ function. The $W$ function
has taken a considerable interest in the literature since it can
be implemented by various means, e.g., \cite{con,ion1,ion2,tom},
and it is sensitive to  the interference in phase space, as we
shall show below. From (\ref{secf5})  we can extract several
analytical facts. For instance, when $(r,m)=(0,0)$ and the value
of $\alpha_2$ is very small, the $W$ function of the first mode
exhibits the well-known shape of the cat-state function, i.e., two
Gaussian bell and interference fringes in-between (we have checked
this fact). This can be easily understood, where--in this
case--the second mode is very close to the vacuum state and hence
(\ref{sec2}) reduces to $|\psi\rangle\simeq
\lambda_\epsilon[|\alpha_1\rangle+\epsilon|-\alpha_1\rangle]\bigotimes|0\rangle$,
i.e. the first mode evolves in the Shr\"{o}dinger-cat state.
Additionally, when $\alpha_2$ increases the negative values in $W$
function gradually decreases  and eventually vanishes showing the
$W$ function of the statistical mixture of coherent states. This
is related to the fact that the interference term in the $W$
function includes the factor $\exp(-2\alpha_2^2)$, which tends to
zero for large values of $\alpha_2$.  On the other hand, when
$|\epsilon|=0, (r,m)\neq (0,0)$, the $W$ function cannot exhibit
neither negative values nor stretching contour in phase space
since ${\rm L}_m(-\kappa)>0$ as $\kappa\geq 0$. In this case, the
behavior of the $W$ function is close to that of the thermal state
for which the peak occurred in the $W$ function is greater than
that of the coherent light.

Now we draw the attention to the general case (see Figs. 4 for
even-type state). For $m=n=0$ we have numerically noted that the
$W$ function exhibits negative values only when
$\alpha_1\geq\alpha_2$. This condition can be analytically
realized as follows.
 From (\ref{secf5})
the interference term in the $W$ function includes $ \cos(\phi
-\frac{4y\Lambda}{\cosh(2r)})$, which is responsible for the
occurrence of
 the negative values in the $W$ function. Assuming that we
choose the  values of the interaction parameters to verify
$\cos(.)=-1$ and  take $(|\epsilon|,x)=(1,0)$ to simplify the
problem. Thus, the $W$ function reduces to
\begin{equation}\label{reply4}
W(0,y)=\frac{4|\lambda_{\epsilon}|^{2}}{\pi \cosh
(2r)}\{\exp[-\frac{2(y^2+\alpha_1^2)}{\cosh(2r)}]
-\exp[-\frac{2(\alpha^2_2+y^{2})}{\cosh(2r)} ]
 \}.
\end{equation}
The $W$ function involves negative values  when $W(0,y)<0$.
Solving this inequality
 yields the above mentioned condition. From Fig. 4(a) one can observe
that the $W$ function has two-Gaussian bell and interference fringes
in between but with negative values smaller than those of the
standard cat states, e.g. \cite{vidi}. These fringes can be
amplified for certain values of  the squeezing parameter (compare
Figs. 4(a) and (b)). It is mentioning worth that the amplification
of the cat states in the parametric down conversion has been
discussed in \cite{filip}. Now we draw the attention to the case in
which the second mode includes  Fock state $|1\rangle$
 (see Fig. 4(c)). From this
figure  the $W$ function exhibits two-Gaussian bell around
$(x,y)=(\pm\alpha_1,0)$ and inverted peak in-between with maximum
negative value. Comparison between Figs. 4(a) and (c) shows that
the existence of  Fock state in the second mode increases the
amounts of the nonclassical effects in the first one. Precisely,
  the interference in phase space in the first mode
can be controlled by the information involved in the second mode.
In this respect the nonclassical effects can be transferred from
one of the modes to the other through the entanglement process.
\begin{figure}
    \includegraphics[width=1.0\linewidth]{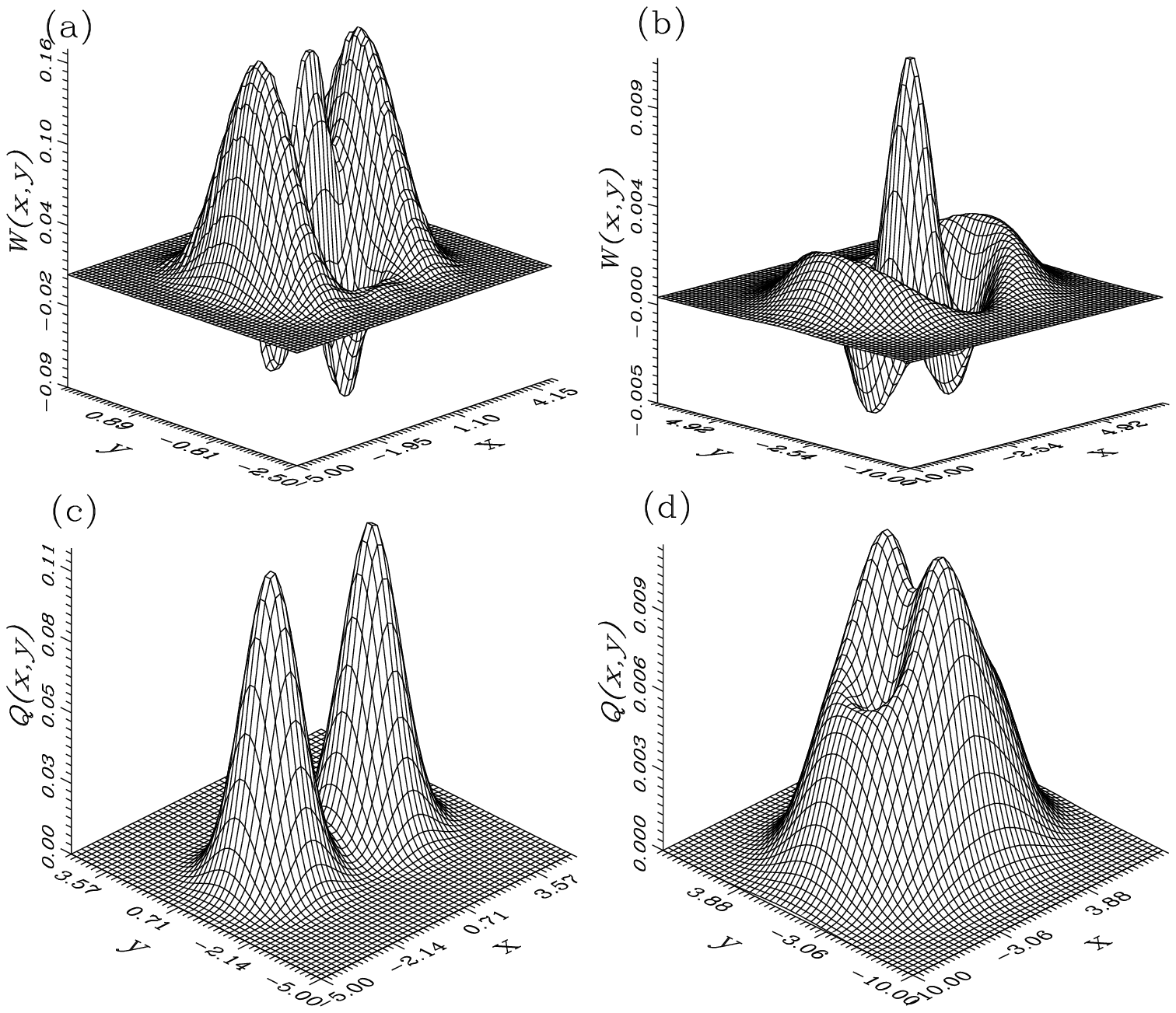}
\caption{ The $W$ function (a), (b) and the $Q$ function (c), (d)
of the STDSN for
$(\alpha_1,\alpha_2,|\epsilon|,\phi,m)=(2,0.9,1,\pi,1)$ with
$r=0.4$ (a),(c) and $1.8$   (b), (d). }
\end{figure}
 Furthermore,
involving the squeezing mechanism in the system smoothes out the
negative values in the $W$ function (compare Figs. 4(c) and (d)),
which vanish  for large values of  $r$, we get back to this point
shortly.  In this case we found that the $W$ behaves quite similar
as that of the thermal light. This is related to the correlation
mechanism in the system.
 Comparison
between Figs. 4(b) and (d) shows that the squeezing mechanism makes
the interference fringes more or less pronounced based on the value
of $m$. Now we draw the attention to Figs. 5 given for $W$ and $Q$
functions, as indicated, for the odd-type states.
 From Figs. 5(a) and (b) $W$ function exhibits negative
values and with more structure compared to those of the even
states (compare these figures  with Figs. 4(a) and (b)). From Fig.
5(b) the negative values  still exist even for  large values of
$r$. We use the expression  "the large values of $r$" when $r\geq
1$. This is inspired by the fact: one of the two-mode quadrature
operators   exhibits maximum squeezing (, i.e., $F\simeq -1$ (c.f.
(\ref{reply12}))) when $r\geq 1$. From Fig. 5(c) one can observe
that $Q$ function exhibits a symmetric two-peak structure, which
is representative to the cat states as well as the statistical
mixture states \cite{vidi}. Moreover, when the value of $r$
increases, i.e. the entanglement between the two modes becomes
stronger, the $Q$ function exhibits a quite similar shape to that
of the thermal light (see Fig. 5(d)), which is a single peak
localized in the phase space origin with contour greater than that
of the vacuum state. In the framework of $W$ ($Q$) function the
first mode exhibits (nonclassical) super-classical light (compare
Figs. 5(b) and (d)). This confirms the fact that: the $W$ function
is more informative than the $Q$ function. Similar conclusions
have been noticed for the case $(|\epsilon|,\phi)=(1,\pi/2)$. We
conclude this part by investigating  the relation between the
occurrence of negative values in the $W$ function and the value of
the squeezing parameter $r$.  To do so we plot Fig. 6 for the $W$
function in terms of $r$ for the same values of the parameters as
in Fig. 4(c) (dashed curve) and Fig. 5(b) (solid curve). The
values of $x$ and $y$ have been chosen as they give maximum
negative values in Fig. 4(c) and Fig. 5(b).
 From Fig. 6 we can obtain a rough information about   the minimal value of  $r$ for which the negative
 values in $W$ function  vanish, e.g. for even-type and odd-type
 states it is
 $r=1$ and $r=2.3$, respectively.  We have checked the behavior of the $W$ function for these values
 and found that the  negative values are negligible. Furthermore, after plotting the $Q$ function for various values of $r$
 (not detailed here) we observed  that the exact minimal  value for even-type state
 is  $r=1.5$. Nevertheless, for the odd-type state we found that the negative values--even they are very small--are
 still observed for all values of $r$.

\begin{figure}
    \includegraphics[width=0.5\linewidth]{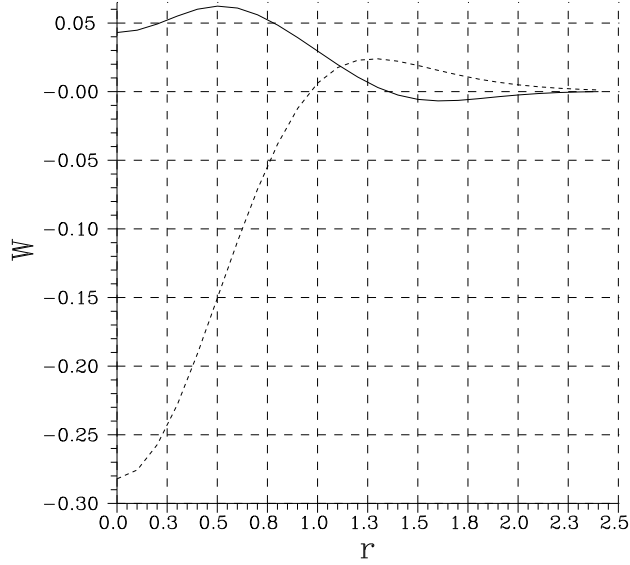}
\caption{ The $W$ function against $r$ for
$(\alpha_1,\alpha_2,|\epsilon|,m)=(2,0.9,1,1)$ with
$(x,y,\phi)=(0,0,0)$ (dashed curve) and $(3,0,\pi)$ (solid curve).
The grid is given to show the bounds of $r$ for which $W=0$.
 }
\end{figure}

Entanglement is a global property of a system.  For a bipartite
pure state it has been proved that there is a unique measure of
the entanglement, which is the von Neumann entropy of the reduced
state of either of the parties \cite{Popescu}. On the other hand,
the purity, which gives information on the mixedness in the
system, can be used to estimate some information on the
entanglement in the system. In this respect, we can mention that
 the purity and the von Neumann entropy can give quite
similar behavior for  the quantum system \cite{bajer}. Also for
the Jaynes-Cummings model it has been shown that the von Neumann
entropy and purity are equivalent \cite{ahmed}. As the purity is
easy to be calculated and can provide some exact information about
the system, we use it here to study the mixedness and/or the
entanglement in the state under consideration. The single-mode
purity can be evaluated via the characteristic function through
the relation:
\begin{equation}\label{purity1}
{\rm Tr}\hat{\rho}_j^2=\frac{1}{\pi}\int |C_{w}(\beta)|^2d^2\beta,
\end{equation}
where $\hat{\rho}_j$ is the density matrix for the mode under
consideration. For pure (mixed) state we have ${\rm
Tr}\hat{\rho}_j^2=1 (<1)$.   From (\ref{secf2}), (\ref{purity1})
and setting $(|\epsilon|,m,n)=(1,0,0)$ we obtain:
\begin{eqnarray}
\begin{array}{lr}
    {\rm
    Tr}\hat{\rho}_1^2=\frac{2|\lambda_\epsilon|^4}{\cosh(2r)}\{1
 +   \mu^2\cos(2\phi)   +\exp\left(-\frac{4\alpha_1^2}{\cosh(2r)}\right)\\
\\
+4\mu\cos\phi
    \exp\left[-\frac{(\alpha_1-\alpha_2)(\alpha_1-\Lambda)}{\cosh(2r)}\right]+
   \mu^2\exp\left(\frac{4\Lambda^2}{\cosh(2r)}\right) \}.\label{pury2}
\end{array}
\end{eqnarray}

 In Figs. 7 we have plotted ${\rm
    Tr}\hat{\rho}_1^2$ for the even-type states.
From Fig. 7(a) it is obvious that for $\alpha_1=0$ or $\alpha_2=0$
the two modes are  disentangled, where  ${\rm
    Tr}\hat{\rho}_1^2=1$. When $\alpha_j$ increases the first mode abruptly tends
    to the partial mixed state (i.e., ${\rm
    Tr}\hat{\rho}_1^2=0.5$), which indicates strong entanglement between the two modes.
    In this case the behavior is quite similar
    to that of the thermal light with mean-photon number
    $\bar{n}=1$, which satisfies the inequality
$\frac{1}{1+\bar{n}}\leq    {\rm
    Tr}\hat{\rho}^2<1$.
    This behavior can be analytically
realized by evaluating  the limiting case
$(\alpha_1,\alpha_2)=(\infty,\infty)$ for the purity
(\ref{pury2}), which gives:
\begin{equation}\label{reply6}
    {\rm
    Tr}\hat{\rho}_1^2=\frac{1}{2\cosh(2r)}.
\end{equation}
It is evident that ${\rm
    Tr}\hat{\rho}_1^2=0.5$ for $r=0$.
    When the squeezing mechanism is involved
    in the system the degree of  mixedness and/or the amount of
    entanglement is increased    and the purity becomes more
structured (compare Fig. 7(a) and (b)). The purity tends to steady
state for large values of $\alpha^{,}s$.
 The value of the
purity at $(\alpha_1,\alpha_2)=(0,0)$ can be easily obtained from
(\ref{pury2}) as:
\begin{equation}\label{reply7}
    {\rm
    Tr}\hat{\rho}_1^2=\frac{1}{\cosh(2r)}.
    \end{equation}
 It is evident when the value of $r$ increases the amount of
mixedness  increases, too.
\begin{figure}
   \includegraphics[width=1.0\linewidth]{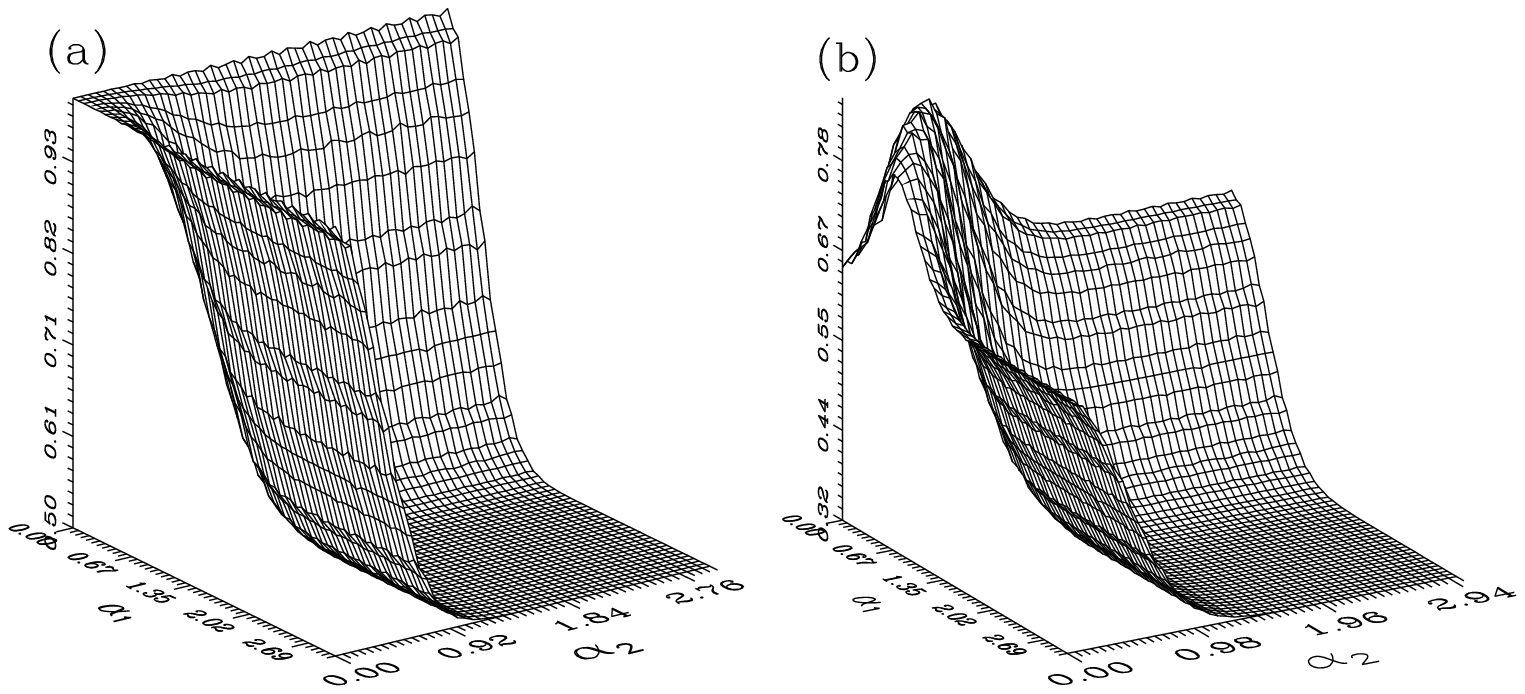}
\caption{ The purity of the first mode  for
$(|\epsilon|,\phi)=(1,0)$ with $r=0$ (a) and $0.5$ (b).}
\end{figure}
From the  expressions (\ref{reply6}) and (\ref{reply7}) one can
realize that the minimal  value of the purity can be achieved for
large $r$; e.g. for $(\alpha_1,\alpha_2,r)=(2,2,5)$ we have ${\rm
    Tr}\hat{\rho}_1^2\simeq 0.006$.

\section{States generation}\label{S:sec6}
In this section we give a generation scheme for  the states
(\ref{sec2}) in the frame work of the trapped ions.  To do so we
consider  a two-level ion of mass $M$ moving in a $2-D$ harmonic
potential of frequency $\omega_x$ in the $x$-direction and
$\omega_y$ in the $y$-direction. Also $\hat{a}\quad
(\hat{a}^{\dagger})$ and $\hat{b}\quad (\hat{b}^{\dagger})$
represent the annihilation (creation) operators for the vibronic
quanta in the $x$- and $y$-directions, respectively. Then the
position operators are given by $\hat{x}=\triangle
x_0(\hat{a}+\hat{a}^{\dagger}), \quad \hat{y}=\triangle
y_0(\hat{b}+\hat{b}^{\dagger})$, where $\triangle x_0=(2\omega_x
M)^{-\frac{1}{2}}, \triangle y_0=(2\omega_y M)^{-\frac{1}{2}}$ the
width of the harmonic ground state. Six beams are used to drive
the interaction with the ion in the cavity; two are propagating in
the $x$-direction detuned by $\pm \omega_x$ from the transition
frequency of the ion. Two are propagating in the $y$-direction
detuned by $\pm \omega_y$ and two are propagating in the $x-y$
plane detuned by $\pm (\omega_x+\omega_y)$. Thus the interaction
Hamiltonian can be written in the form:

\begin{equation}\label{trao2}
\hat{H}_{int}=-(\hat{\mu}.\hat{E}^{-}\hat{\sigma}_-+h.c),
\end{equation}
where
\begin{eqnarray} \label{trao3}
\begin{array}{lr}
\hat{E}^{-}=E_1\exp[i((\omega_0-\omega_x)t-k_1x+\vartheta_1)]
+ E_2\exp[i((\omega_0+\omega_x)t-k_2x+\vartheta_2)]\\
\\
+ E_3\exp[i((\omega_0-\omega_y)t-k_3y+\vartheta_3)] +
E_4\exp[i((\omega_0+\omega_y)t-k_4y+\vartheta_4)]
\\
\\
+ E_5\exp[i((\omega_0-\omega_x-\omega_y)t-k_5x-k'_5y+\vartheta_5)]
+
E_6\exp[i((\omega_0+\omega_x+\omega_y)t-k_6x-k'_6y+\vartheta_6)],
\end{array}
\end{eqnarray}
with $\hat{\sigma}_\pm$ are the Puali spin operators,  $E_j, k_j,
\vartheta_j$ are the amplitudes, wave vectors, phases of the
driving modes, and $\omega_0$ is the ionic transition frequency.
 Using the operator forms for $x,y$ and providing that, the
field is resonant with one of the vibronic side - bands, then the
ion-field interaction can be described by a nonlinear
Jaynes-Cummings model \cite{vogel}. In the interaction picture and
in the Lamb-Dicke limit, it is sufficient  to keep the first few
terms. Thus we have the following effective Hamiltonian:

\begin{eqnarray} \label{trao3}
\begin{array}{lr}
\hat{H}_{int}=\hat{H}_{1}+\hat{H}_{2},\\
\\
\hat{H}_{1}=-(g_1\hat{a}^{\dagger}+g_2\hat{a}+
g_3\hat{b}^{\dagger}+g_4\hat{b})\hat{\sigma}_-+h.c,\\
\\
\hat{H}_{2}=-(g_5\hat{a}\hat{b}+g_6\hat{a}^{\dagger}\hat{b}^{\dagger})
\hat{\sigma}_++h.c
\end{array}
\end{eqnarray}
where
\begin{eqnarray} \label{trao4}
\begin{array}{lr}
g_j=i\Omega_j\eta_j\exp(i\vartheta_j-\frac{1}{2}\eta_j^2),\quad j=1,2,3,4\\
\\
g_j=-\Omega_j\eta_j\eta'_j\exp[i\vartheta_j-\frac{1}{2}(\eta_j^2+\eta_j^{'2})],\quad
j=5,6,
\end{array}
\end{eqnarray}
with $\Omega_j=\mu E_j$ and $\eta_j^2=k_j^2(\triangle x_0)^2,
j=1,2,5,6,\quad \eta_j^2=k_j^2(\triangle y_0)^2, j=3,4,5,6$ and
stand for the Rabi frequency and the Lamb-Dicke parameter. The
motional and internal dynamics can be described in the last
Hamiltonian by adding other interactions as discussed in
\cite{stein} to end up with

\begin{eqnarray} \label{trao5}
\begin{array}{lr}
\hat{\bar{H}}_{1}=-[(g_1+g_2^{*})\hat{a}^{\dagger}+(g_1^{*}+g_2)\hat{a}+
(g_3+g_4^{*})\hat{b}^{\dagger}+(g_3^{*}+g_4)\hat{b}]
(\hat{\sigma}_-+\hat{\sigma}_+),\\
\\
\hat{\bar{H}}_{2}=-[(g_5+g_6^{*})\hat{a}\hat{b}+
(g_5^{*}+g_6)\hat{a}^{\dagger}\hat{b}^{\dagger}]
(\hat{\sigma}_-+\hat{\sigma}_+).
\end{array}
\end{eqnarray}
Under this Hamiltonian any particle prepared in the state
$\frac{1}{\sqrt{2}}(|e\rangle+|g\rangle)$ will stay in this state
and the dynamics is reduced to that of the motional degrees of
freedom only. Now assuming that the system is initially prepared
in the following state:
\begin{equation}\label{trapo}
    |\Psi(0)\rangle=(|e\rangle+|g\rangle)|n,m\rangle,
\end{equation}
where $|e\rangle, |g\rangle$ denote the excited and the ground state
of the ion. Also we have dropped the normalization constant in
(\ref{trapo}) since it has no effect in the following calculations.
It is worth mentioning that the Fock state $|n\rangle$ can be
prepared with very high efficiency according to the recent
experiments \cite{Leib}. We proceed, by applying the Hamiltonian
$\hat{\bar{H}}_{2}$ on the state (\ref{trapo}) for a duration time
$\tau_1$ we get
\begin{equation}\label{trao6}
    |\Psi_1\rangle=\exp(-i\hat{\bar{H}}_{2} \tau_1)|\Psi(0)\rangle=
    \hat{S}(r)|n,m\rangle(|e\rangle+|g\rangle).
\end{equation}
Then we apply $\hat{\bar{H}}_{1}$ for a duration $\tau_2$ to get:
\begin{equation}\label{trao7}
    |\Psi_2\rangle=\exp(-i\hat{\bar{H}}_{1}
\tau_2 )|\Psi_1\rangle=D(\alpha_1,\alpha_2)
    \hat{S}(r)|n,m\rangle(|e\rangle+|g\rangle),
\end{equation}
where $r=i(g_5+g_6^*)\tau_1,\quad \alpha_1=-i(g_1+g_2^*)\tau_2$
and $\alpha_2=-i(g_3+g_4^*)\tau_2$. We choose the polarization in
the quantized field so that it affects the excited state only
\cite{Leib} and apply the Hamiltonian $\hat{\bar{H}}_{1}$ for a
duration $\tau_3$ we arrive at
\begin{equation}\label{trao8}
    |\Psi_3\rangle=\exp(-i\hat{\bar{H}}_{1}
    \tau_3)|\Psi_2\rangle=[(D(\beta_1,\beta_2)|e\rangle+D(\alpha_1,\alpha_2)|g\rangle]
    \hat{S}(r)|n,m\rangle,
\end{equation}
where $\beta_1=\alpha_1-i(g_1+g_2^*)\tau_3$ and
$\beta_2=\alpha_2-i(g_3+g_4^*)\tau_3$. After that we apply a
carrier pulse of Rabi frequency $\Omega_0$, whose evolution
operator is
\begin{equation}\label{trao9}
\hat{U}(t)=\cos(\Omega_0 t) (|e\rangle\langle e|+|g\rangle\langle
g|)-i\sin(\Omega_0 t) (\exp(i\theta)|e\rangle\langle
g|+\exp(-i\theta)|g\rangle\langle e|),
 \end{equation}
to the state $|\Psi_3\rangle$ to get

\begin{eqnarray} \label{trao10}
\begin{array}{lr}
|\Psi_4\rangle=\hat{U}(\tau_4)|\Psi_3\rangle=[D(\beta_1,\beta_2)\cos(\Omega_0
\tau_4)-i\exp(i\theta)\sin(\Omega_0 \tau_4)D(\alpha_1,\alpha_2)]
\hat{S}(r)|n,m\rangle |e\rangle\\
\\
+\left[D(\alpha_1,\alpha_2)\cos(\Omega_0 \tau_4)
-i\exp(-i\theta)\sin(\Omega_{0} \tau_{4})
D(\beta_1,\beta_2)\right] \hat{S}(r)|n,m\rangle |g\rangle.
\end{array}
\end{eqnarray}
Then detecting the particle in either of its states gives  the
states (\ref{sec2}), where
 we can choose $\beta_1=-\alpha_1,
\beta_2=-\alpha_2$.  Throughout the investigation of this paper we
have considered that the parameters $\alpha_j$ and $r$ are real.
This can be achieved  in the above equations  by simply setting,
e.g., $ \vartheta_j=0$ or $\pi, j=1,..,4$ and $ \vartheta_5=
\vartheta_6=\pi/2$.

\section{Conclusion}\label{S:sec7}
The superposition principle is in the heart of  quantum mechanics,
which can produce new states having nonclassical effects greater
than those attributed to the components. In this article we have
studied the quantum properties for a new class of states, namely,
superposition of squeezed displaced two-mode number states.
Particular attention has been given to the two-mode vacuum and
single-photon states of this class. These states include two
mechanisms: interference in phase space and entanglement between
the two modes of the system.  We have studied the second-order
correlation function, the Cauchy-Schwartz inequality, the
quadrature squeezing, the quasiprobability distribution functions
and the purity. We have shown that the system can exhibit
sub-Piossonian statistics even if the mode under consideration is
in the vacuum state. This reflects the role of entanglement in the
system. The deviation from the classical Cauchy-Schwartz
inequality has been investigated showing that the photons are more
strongly correlated than it is allowed classically.
 For certain values of $\epsilon$ the
system can exhibit squeezing  provided that the values of
$\alpha_1$ and $\alpha_2$ are small. From the $W$ function it has
been shown that the single-mode state resulting from this class
can behave as a thermal state as a result of the correlation
process. Also for $m=n=0$ the $W$ function of the first mode
provides negative values only when $\alpha_1\geq\alpha_2$. The
interference in phase space of one of the subsystem can be
controlled by the information involved in the other subsystem.
Additionally, the squeezing mechanism can make the interference
fringes more or less pronounced. The $W$ function is more
informative than the $Q$ function in the description of the
quantum systems. For the purity it has been shown when the values
of $\alpha^{,}s$ increase the mode under consideration abruptly
tends
    to the partially mixed state. The amount of entanglement in the system is increased when the
value of  $r$ is increased, too. Also we have discussed how this
class of states can be generated by means of trapped ions and pulses
for appropriate durations.

\section*{Appendix}
 In this Appendix we prove that the $W$ and $Q$ functions (\ref{secf5})
 are normalized. Precisely, we would like to prove the followings:

\begin{equation}\label{rrefr1}
  \int W(z)d^2z=1,\quad \int Q(z)d^2z=1.
\end{equation}
  To do
so we  use the generating function technique. We focus the
attention on the $Q$ function only, where the $W$ function can be
similarly treated.  Moreover, we evaluate  the integration for one
of the interference terms in the $Q$ function, which we denote
$I_m$ and has the form:

\begin{equation}\label{appen1}
    I_m=\frac{1}{\pi C^{2m+2}_r}\int
\exp[-\frac{1}{C_r^2}(\alpha_1^2+2t^2_2+|z|^{2})] \exp\left(-i\phi
-\frac{(z-z^*)\Lambda}{C^2_r}\right) {\rm L}_m(h')d^2z
\end{equation}
Multiply both sides of (\ref{appen1}) by $t^m$, hence sum over
index $m$ and use the identity (\ref{reply30}) we obtain:
\begin{equation}
 \sum\limits_{m=0}^{\infty}t^mI_m=\frac{\exp(-i\phi)}{\pi(1-t')C_r^2} \exp
\left[-\frac{(\alpha_1^2+2t_2^2)}{C_r^2}\right]
 \int \exp \left[-\frac{|z|^2}{C_r^2}-\frac{\Lambda (z-z^*)}{C_r^2}+\frac{h' t'}{t'-1}\right] d ^{2}z
,\label{refr2}
\end{equation}
where $t'=t/C_r^2$. Invoking the value of $h'$ from (\ref{scf5})
into (\ref{refr2}) and apply the identity (\ref{reply31}) we
arrive at:

\begin{eqnarray}
\begin{array}{lr}
 \sum\limits_{m=0}^{\infty}t^mI_m=\frac{\exp(-i\phi)}{1-t}
\exp\left[-\frac{(\alpha_1^2+2t_2^2)}{C_r^2}+\frac{t'(\alpha_2C_r+t_2)^2}{(t'-1)C_r^2}\right]\\
\\
\times
 \exp
\{-\frac{[S_r(2\alpha_2C_r+\alpha_1S_r)t'-\Lambda(t'-1)]^2}{(t'-1)(t-1)C_r^2}\}.\label{referr3}
\end{array}
\end{eqnarray}
The exponent in the above equation can be rewritten in terms of
the parameter $t$ as:
\begin{eqnarray}
\begin{array}{lr}
 \sum\limits_{m=0}^{\infty}t^mI_m=\frac{\exp(-i\phi)}{1-t}
\exp\left[-\frac{(\alpha_1^2+2t_2^2+\Lambda^2)}{C_r^2}-\frac{[2\alpha_2C_r+\alpha_1 S_r+S_r \Lambda]^2t}{C_r^4(t-1)}\right]\\
\\
=\exp(-i\phi)
\exp\left[-\frac{(\alpha_1^2+2t_2^2+\Lambda^2)}{C_r^2}\right]
\sum\limits_{m=0}^{\infty}t^m {\rm L}_m
\left[\frac{(2\alpha_2C_r+\alpha_1 S_r+S_r
\Lambda)^2}{C_r^4}\right].\label{referr4}
\end{array}
\end{eqnarray}
The transition from the first line to the second one has been done
by means of the  identity (\ref{reply30}). Now the value of the
required integral is

\begin{equation}\label{referr5}
I_m=\exp(-i\phi)
\exp\left[-\frac{(\alpha_1^2+2t_2^2+\Lambda^2)}{C_r^2}\right] {\rm
L}_m \left[\frac{(2\alpha_2C_r+\alpha_1 S_r+S_r
\Lambda)^2}{C_r^4}\right].
\end{equation}
Through minor treatments one can easily prove:
\begin{equation}\label{referr6}
\frac{(\alpha_1^2+2t_2^2+\Lambda^2)}{C_r^2}=2(t_1^2+t_2^2),\quad
\frac{(2\alpha_2C_r+\alpha_1 S_r+S_r \Lambda)^2}{C_r^4}=4t_2^2.
\end{equation}
Therefore, the quantity $I_m$ takes the form:
\begin{equation}\label{referr7}
I_m=\exp(-i\phi)\exp[-2(t_1^2+t_2^2)]
 {\rm L}_m (4t_2^2).
\end{equation}
The value of the integration of the second interference term is
just the complex conjugate of (\ref{referr7}). Similar procedures
lead to the followings:

\begin{equation}\label{referr8}
\frac{\pi} {C^{2m+2}_r} \int  \exp[-\frac{|z-\alpha_1|^2}{C_r^2}]
{\rm L}_m[-|z-\alpha_1|^2\tanh^2 r]d^2z= \frac{\pi}{
C^{2m+2}_r}\int \exp[-\frac{|z+\alpha_1|^2}{C_r^2}] {\rm
L}_m[-|z+\alpha_1|^2\tanh^2 r]d^2z=1.
\end{equation}
From these  results we conclude:
\begin{equation}\label{referr9}
  \int
  Q(z)d^2z=|\lambda_{\epsilon}|^{-2}|\lambda_{\epsilon}|^{2}=1.
\end{equation}
Using procedures similar to those given above one can prove that
the $W$ function is normalized (we have checked it).

\section*{ Acknowledgement}

The authors would like to thank Professor Margarita Man'ko for the
critical reading of the manuscript and for many suggestions, which
have improved this paper.

\section*{References}

\end{document}